\documentclass[12pt,preprint]{aastex}
\usepackage[dvips]{color}

\newcommand{\dpp}[2]{\frac{\partial #1}{\partial #2}}
\newcommand{\ddt}[1]{\dpp{#1}{t}}

\newcommand{\ddx}[1]{\dpp{#1}{x}}

\newcommand{\ddz}[1]{\dpp{#1}{z}}

\newcommand{\mi}[1]{\mbox{\boldmath$#1$}}
\newcommand{\Nbv}{\mathcal{N}^{\,2}}

\shorttitle{MHD waves in inclined magnetic field}
\shortauthors{Parchevsky \& Kosovichev}

\begin{document}
\title{Numerical simulation of excitation and propagation of helioseismic
MHD waves: Effects of inclined magnetic field}

\author{Parchevsky, K.V., Kosovichev, A.G.}
\affil{Stanford University, HEPL, Stanford CA 94305, USA}
\email{kparchevsky@solar.stanford.edu}

\begin{abstract}
Investigation of propagation, conversion, and scattering of MHD
waves in the Sun is very important for understanding the mechanisms
of observed oscillations and waves in sunspots and active regions.
We have developed 3D linear MHD numerical model to investigate
influence of the magnetic field on excitation and properties of the
MHD waves. The results show that the magnetic field can
substantially change the properties of the surface gravity waves
($f$-mode), but their influence on the acoustic-type waves
($p$-modes) is rather moderate. Comparison our simulations with the
time-distance helioseismology results from SOHO/MDI shows that the
travel time variations caused by the inclined magnetic field do not
exceed 25\% of the observed amplitude even for strong fields of
1400--1900 G. This can be an indication that other effects (e.g.
background flows and non-uniform distribution of magnetic field) can
contribute to the observed travel time variations. The travel time
variations caused by the wave interaction with magnetic field are in
phase with the observations for strong fields of 1400--1900 G if
Doppler velocities are taken at the height of 300 km above the
photosphere where plasma parameter $\beta\ll1$. The simulations show
that the travel times only weakly depend on the height of velocity
observation. For the photospheric level the travel times are
systematically smaller on approximately 0.12 min then for the hight
of 300 km above the photosphere for all studied ranges of the
magnetic field strength and inclination angles. The numerical MHD
wave modeling and new data from the HMI instrument of the Solar
Dynamics Observatory will substantially advance our knowledge of the
wave interaction with strong magnetic fields on the Sun and improve
the local helioseismology diagnostics.

\end{abstract}


\section{Introduction}
Local helioseismology has provided important results about the
structures and dynamics of the solar plasma below the visible
surface of the Sun, associated with sunspots and active regions
\citep[e.g.][]{Duvall1996,Kosovichev1996,Kosovichev2000,Zhao2001,Haber2000,Komm2008}.
The helioseismic inferences help us to understand the complicated
processes of the origin solar magnetic structures, formation and
evolution of sunspots and active regions. These studies are based on
measurements and inversions of variations of acoustic travel times
and oscillation frequencies in the areas occupied by magnetic field
and around them.

There are several factors that may cause the observed variations of
the oscillation properties, and it is very important to
painstakingly investigate their effects for improving the
reliability of the helioseismic inference \citep{Bogdan2000}. Such
studies are carried out both observationally by doing various
experiments with the data analysis procedure, e.g. by masking the
regions of strong field, doing "double-skip" experiments etc.
\citep[e.g.][]{Zhao2006}, and theoretically by simulating wave
propagation in various conditions of the solar convection zone and
calculating how these conditions affect the helioseismic
observables, such as the oscillation power spectrum and acoustic
travel times \citep{Georgobiani2007,Zhao2007}. We emphasize that for
correct interpretation of helioseismic results the theoretical
modeling must include calculations of the actual observables, taking
into account all important aspects of the data measurement
procedure, such as data filtering and averaging, and geometrical
factors \citep{Nigam2007}. Unfortunately, in many theoretical
studies the actual helioseismic measurement procedure is not
modeled, and this may lead to incorrect conclusions about the role
of various factors in the helioseismic results.

In general, the main factors causing variations in helioseismic
travel times in solar magnetic regions, can be divided in two types:
direct and indirect. The direct effects are due to the additional
magnetic restoring force, which changes the wave speed and may
transform acoustic waves into different types of MHD waves. The
indirect effects are due to the changes in the convective and
thermodynamic properties in magnetic regions. These include
depth-dependent variations of temperature and density, large-scale
flows and changes in wave source distribution and strength. Both
direct and indirect effects may be present in the observed
travel-time and frequency variations and cannot be easily
disentangled by data analyses causing confusions and
misinterpretations. Thus, it is very important to investigate the
various aspects by numerically modeling the individual factors
separately.

In particular, we have investigated the effects of the suppressed
excitation of acoustic waves in sunspot regions, where strong
magnetic field inhibits convective motions, which are the primary
source of solar waves \citep{Parchevsky2007a}. The results showed
that the suppression of acoustic sources may explain most of the
observed deficit of acoustic power in sunspot regions
\citep{Parchevsky2007b}, and also cause systematic shifts in the
travel-time measurements. However, these shifts are significantly
smaller than the observed variations of the travel times and and
also have the opposite sign \citep{Parchevsky2008}, and, thus,
cannot affect the basic conclusions on the sunspot sound-speed
structure, contrary to previous suggestions
\citep[e.g.][]{Rajaguru2006}.

The goal of this paper is to model the excitation and propagation of
helioseismic waves (both f- and p-modes) in the presence of inclined
magnetic field and investigate the importance of the inclined field
in the time-distance helioseismology measurements by carefully
modeling the measurement procedure and comparing with the
observational results, obtained by \citet{Zhao2006}. The issue of
the influence of the inclined magnetic field was raised by
\citet{Schunker2005}, who found that the phase shift of the signal
in the penumbra of a sunspot, measured by the acoustic holography
technique varies with the sunspot position on the disk. They
attributed this to the variations of the angle between the inclined
magnetic field of the penumbra and the line-of-sight. They suggested
that the variations of the phase shift may affect the inferences of
the sound-speed distribution below sunspots, inferred by
time-distance helioseismology. However, \citet{Zhao2006} repeated
the analysis of the same sunspot by the time-distance technique, and
found substantially smaller variations with the position on the
disk, and no significant effect on the wave-speed profile. They also
found that the variations due to the inclination angle exist only
for the wave measurements using the Doppler-shift signal, and that
the variations are absent when the travel times are measured from
the simultaneous intensity observations from SOHO/MDI. This result
indicates that the observed variations with the inclination angle of
the Doppler-shift measurements are likely to be related to changes
of the ratio between the vertical and horizontal components of the
displacement vector of the solar oscillations in the penumbra, and
not the wave transformation or other effects, which could affect the
modal structure of the oscillations. The solar oscillation theory
predicts that the ratio between the vertical and horizontal
components mostly depends on the surface boundary conditions
\citep[e.g.][]{Unno1989}. In the sunspot umbra, the boundary
conditions may change due to the inclined magnetic field or/and near
surface flows, the Evershed effect, which is observed directly in
the Doppler-shift data and shows a significant center-to-limb
variation.

In this paper, we present the results of numerical modeling of the
inclined magnetic field on the time-distance helioseismology
measurements by isolating this affect in a simple magnetic
configuration, and show that only 25\% of travel time variations
measured by \citet{Zhao2006} can be explained by a direct influence
of the inclined magnetic field on acoustic waves. In Sec. 2, we
present the governing equation and describe the numerical method of
3D modeling of helioseismic MHD waves. In Sec.~3, we present the
code verification results comparing the numerical results with
analytical solutions for simple cases. In Sec.~4, we present the
simulation results of the wave propagation in regions with inclined
magnetic field, calculation of the center-to-limb variations of the
time-distance helioseismology measurements and comparison with the
observational results.

\section{Numerical Model}

\subsection{Governing equations and numerical scheme}

Propagation of MHD waves inside the Sun in the presence of magnetic
field is described by the following system of linearized MHD
equations:
\begin{equation}\label{Eq:MHD_3D}
\begin{array}{l}
\displaystyle \ddt{\rho'} + \nabla\cdot \mi{m}'=0,\\[12pt]
\displaystyle \ddt{\mi{m}'} + \nabla p' - \frac{1}{4\pi\rho_0}
[(\nabla\times \mi{B}_0)\times\mi{B}' + (\nabla\times
\mi{B}')\times\mi{B}_0] = \mi{g}_0\rho' + \mi{S},\\[12pt]
\displaystyle \ddt{p'} + c_{s0}^2\nabla\cdot \mi{m}' +
c_{s0}^2\frac{\Nbv_0}{g_0}m_z = 0,
\end{array}
\end{equation}
where $\mi{m}'=\rho_0\mi{v}'$ is the momentum perturbation,
$\mi{v}'$, $\rho'$, $p'$, and $\mi{B}'$ are the velocity, density,
pressure, and magnetic field perturbations respectively, $\mi{S}$ is
the wave source function. The quantities with subscript 0, such as
gravity $\mi{g}_0$, sound speed $c_{s0}$, and Br\"unt-V\"ais\"al\"a
frequency $\mathcal{N}_0$ correspond to the background model. The
spatial and temporal behavior of the wave source is modeled by
function $f(x,y,z,t)$:
\begin{equation}
f(x,y,z,t)=\left\{
\begin{array}{ll}
\displaystyle
A\left[1-\frac{r^2}{R_{src}^2}\right]^2\left(1-2\tau^2\right)e^{-\tau^2} & \mbox{if } r\leq R_{src}\\
\displaystyle 0 & \mbox{if } r>R_{src},
\end{array}
\right.
\end{equation}
where $R_{src}$ is the source radius,
$r=\sqrt{(x-x_{src})^2+(y-y_{src})^2+(z-z_{src})^2}$ is the distance
from the source center, $\tau$ is given by equation
\begin{equation}\label{Eq:MHD_3Dsource}
\tau=\frac{\omega (t-t_0)}{2} - \pi, \qquad t_0\leq t\leq
t_0+\frac{4\pi}{\omega},
\end{equation}
where $\omega$ is the central source frequency, $t_0$ is the moment
of the source initiation. In our simulations we used sources of two
types: source of vertical force $\mi{S}=(0,0,f)^T$ and pressure
source $\mi{S}=\nabla f$. Superposition of such randomly distributed
sources describes very well the observed solar oscillation spectrum
\citep{Parchevsky2007a}.

For numerical solution of equations (\ref{Eq:MHD_3D}) a
semi-discrete finite difference scheme of high order was used. At
the top and bottom boundaries non-reflective boundary conditions
based on the perfectly matched layer (PML) technique were set. We
used the standard solar model S \citep{Christensen-Dalsgaard1996}
with a smoothly joined model of the chromosphere of
\citet{Vernazza1976}, as the background model. The background model
was modified near the photosphere to make it convectively stable.
Details of numerical realization of the code and the background
model can be found in \citet{Parchevsky2007a}.

\subsection{Code verification for different types of MHD waves}
To verify the code and estimate the total error of the method we
compare numerical results with simple analytical solutions. We
assume that all quantities depend on time $t$ and one spatial
coordinate $x$ only, and consider the following linearized adiabatic
1D system of the MHD equations in Cartesian coordinates for a
uniform background model
\begin{equation}\label{Eq:MHD_1DxEq}
\begin{array}{l}
\displaystyle \ddt{\rho}+\bar{\rho}\ddx{u}=0,\\[12pt]
\displaystyle \ddt{u} + \frac{1}{\bar\rho}\ddx{p} +
\frac{\bar{B}_z}{4\pi\bar{\rho}}\ddz{B_z}=0,\\[12pt]
\displaystyle \ddt{v} - \frac{\bar{B}_x}{4\pi\bar{\rho}}\ddx{B_y}=0,\\[12pt]
\displaystyle \ddt{w} - \frac{\bar{B}_x}{4\pi\bar{\rho}}\ddx{B_z}= 0,\\[12pt]
\displaystyle \ddt{B_y}-\bar{B}_x\ddx{v} =0,\\[12pt]
\displaystyle \ddt{B_z} + \bar{B}_z\ddx{u} - \bar{B}_x\ddx{w}
=0,\\[12pt]
\displaystyle \ddt{p}+\bar c_s^2\bar\rho\ddx{u} = 0,
\end{array}
\end{equation}
where $\bar c_s$ is the sound speed in the background model, $p$,
$\rho$ are the pressure and density, $u$, $v$, and $w$ are the
velocity components, $B_x$, $B_y$, $B_z$ are the components of the
magnetic field respectively. The quantities related to the
background model are marked here by overbar. By rotating of the
coordinate frame around OZ axis we set $\bar{B}_y=0$. Equations
(\ref{Eq:MHD_1DxEq}) can be rewritten in matrix notations
\begin{equation}\label{Eq:MHD_1DxMat}
\begin{array}{l}
\displaystyle \ddt{\mi{U}}+\mi{A}\ddx{\mi{U}}=0,
\end{array}
\end{equation}
where $\mi{U}=(\rho,u,v,w,B_y,B_z,p)^T$. We seek a solution of
equations (\ref{Eq:MHD_1DxMat}) in infinite interval $-\infty < x <
\infty$ in the form of plane waves
\begin{equation}\label{Eq:MHD_1DxPlnWaveSolution}
\mi{U}=\hat{\mi{U}}\exp\left[i(kx-\omega t)\right]
\end{equation}
Substituting equation (\ref{Eq:MHD_1DxPlnWaveSolution}) into system
(\ref{Eq:MHD_1DxMat}) we obtain an eigenvalue problem for amplitude
$\hat{\mi{U}}$
\begin{equation}
\mi{A}\hat{\mi{U}} = \frac{\omega}{k}\hat{\mi{U}},
\end{equation}
where $V=\omega/k$ is the phase speed of the wave. Thus, the
amplitudes of various MHD quantities are the components of
eigenvectors of matrix $\mi{A}$, and eigenvalues of this matrix
represent phase velocities of the corresponded waves. From this we
calculate the amplitudes of the entropy, Alfven, slow, and fast MHD
waves:
\begin{equation}\label{Eq:MHD_1DxAmplitudes}
\begin{array}{ccccccc}
\hat{\mi{U}}_s=(\bar\rho,&0,&0,&0,&0,&0,&0)^T\\[5pt]
\hat{\mi{U}}_{\mp A}=(0,&0,&\pm V_A,&0,&\bar{B}_x,&0,&0)^T\\[5pt]
\hat{\mi{U}}_{\mp S}=(\bar\rho,&\mp V_S,&0,
&\displaystyle\mp\frac{c_A^2V_S\sin\theta\cos\theta}{V_A^2-V_S^2},&0,
&\displaystyle-\frac{B_0V_S^2\sin\theta}{V_A^2-V_S^2},&c_s^2\bar\rho)^T\\[12pt]
\hat{\mi{U}}_{\mp F}=(\bar\rho,&\mp V_F,&0,
&\displaystyle\mp\frac{c_A^2V_F\sin\theta\cos\theta}{V_A^2-V_F^2},&0,
&\displaystyle-\frac{B_0V_F^2\sin\theta}{V_A^2-V_F^2},&c_s^2\bar\rho)^T.
\end{array}
\end{equation}
The phase velocities of these waves are $0$, $\mp V_A$, $\mp V_S$,
and $\mp V_F$ respectively, where
\begin{equation}\label{Eq:MHD_1DxEigs}
\begin{array}{l}
V_A=c_A\cos\theta,\\
\displaystyle
V_S=\frac{1}{\sqrt{2}} \sqrt{c_s^2+c_A^2-\sqrt{c_s^4+c_A^4-2c_s^2c_A^2\cos2\theta}},\\[12pt]
\displaystyle V_F=\frac{1}{\sqrt{2}}
\sqrt{c_s^2+c_A^2+\sqrt{c_s^4+c_A^4-2c_s^2c_A^2\cos2\theta}}.
\end{array}
\end{equation}
Here $c_A=\bar B/\sqrt{4\pi\bar\rho}$ is the Alfven speed, $\theta$
is the angle between the wave vector and the background magnetic
field ($\bar B_x=\bar B\cos\theta,\; \bar B_z=\bar B\sin\theta$).
Equations (\ref{Eq:MHD_1DxPlnWaveSolution}),
(\ref{Eq:MHD_1DxAmplitudes}), and (\ref{Eq:MHD_1DxEigs}) give us an
analytical solution to 1D system of MHD equations
(\ref{Eq:MHD_1DxEq}), which includes all types of MHD waves. Initial
conditions for these waves are obtained by setting $t=0$ in equation
(\ref{Eq:MHD_1DxPlnWaveSolution}).

\begin{figure}
\epsscale{0.8}\plotone{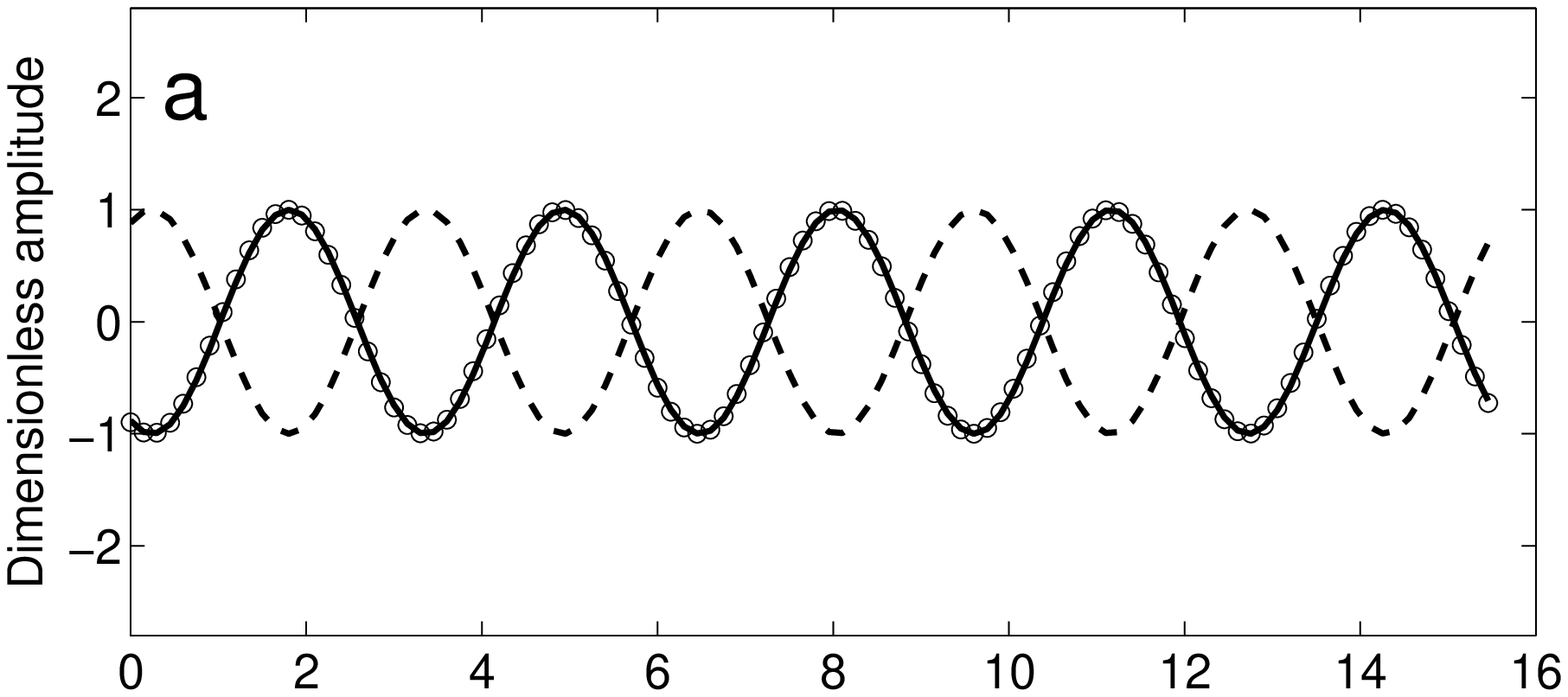}
\epsscale{0.8}\plotone{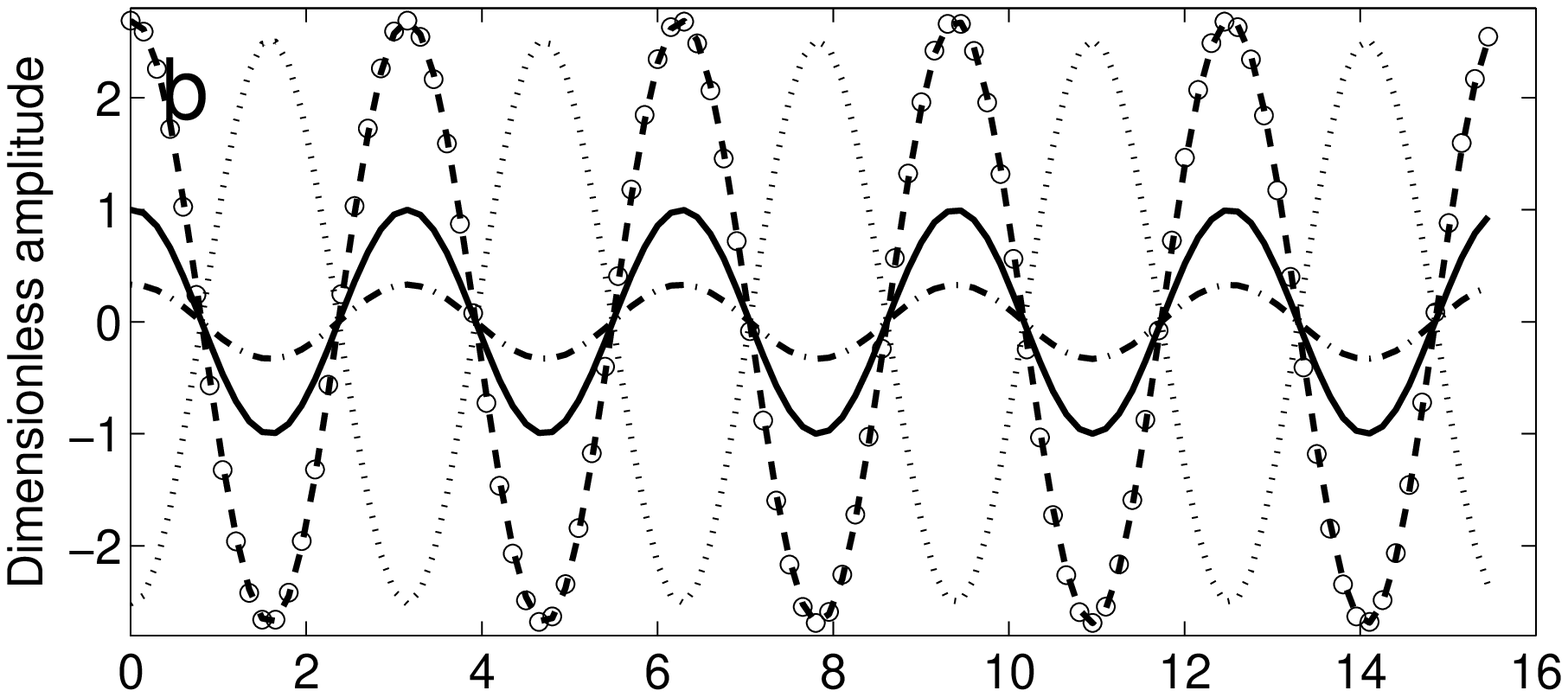}
\epsscale{0.8}\plotone{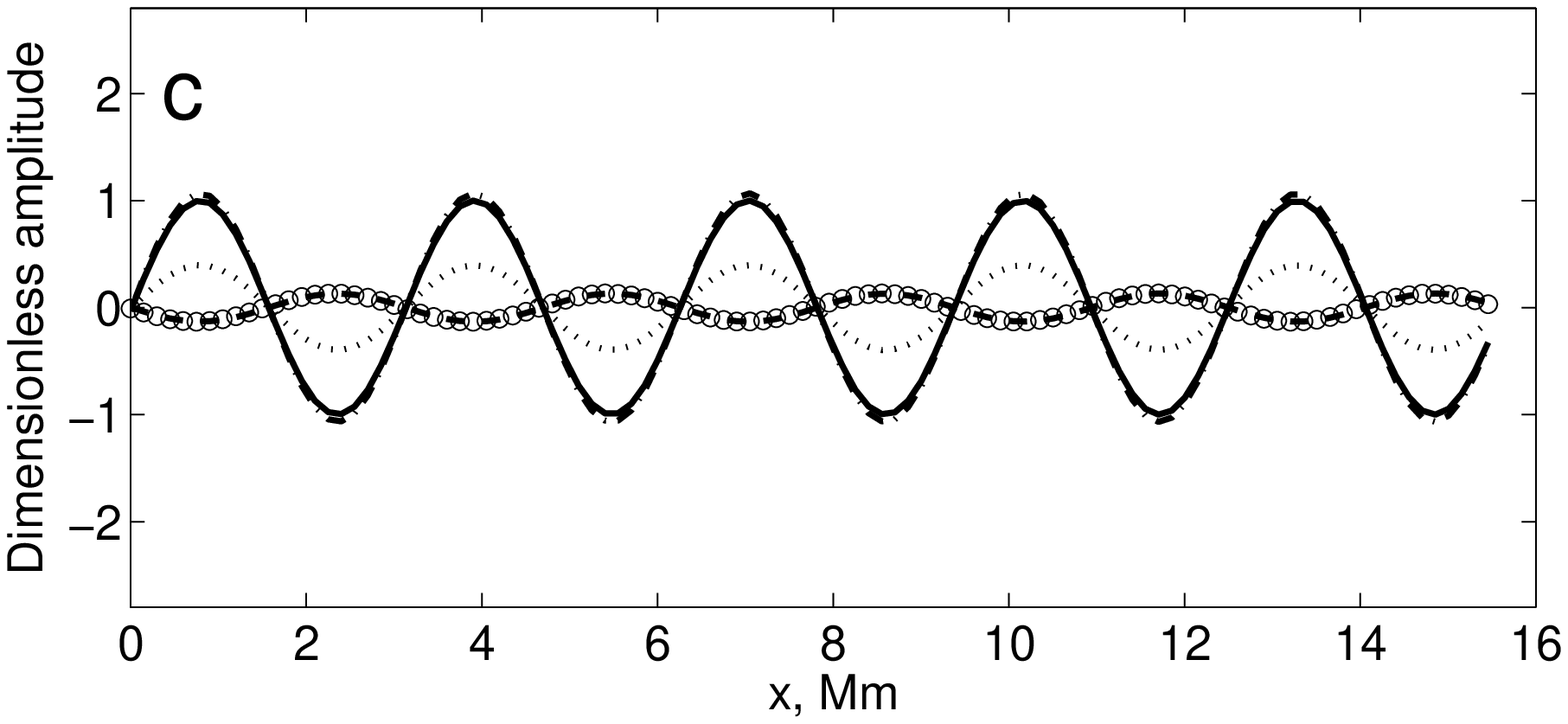}
\caption{\label{Fig:1Dtest_PlnWaves}
Comparison of analytic (cirves) and numerical (markers) solutions
for plain Alfven (a), slow MHD (b), and fast MHD (c) waves
respectively. For Alfven wave only $v$ and $B_y$ (marked by solid
and dashed curves respectively) are non-zero. Circles mark an
analytical solution for $v$. Solid, dash-dotted, dashed, and dotted
curves in panels (b) and (c) correspond to $\rho$, $u$, $w$, and
$B_z$ respectively. Analytical solution for $w$ is shown by
circles.}
\end{figure}

For testing the numerical simulations we use our 3D code with
initial conditions depending only on $x$ variable. The calculations
are carried out in the Cartesian geometry in the domain of
15.46~Mm$\times$15.46~Mm$\times$3.05~Mm with the numbers of grid
points: $N_x=N_y=104,\; N_z=71$. All boundary conditions are chosen
periodic simulating an infinite spatial domain. Wave vector
$k=10\pi/(N_x+1)\Delta x$ is chosen in a way to match the periodic
boundary conditions. In dimensionless variables
$\tilde\rho=\rho/\bar\rho$, $\tilde p=p/\bar\rho \bar c_s^2$,
$\tilde{\mi{v}}=\mi{v}/\bar c_s$, and $\tilde B^2=B^2/4\pi\bar\rho
\bar c_s^2$ parameters of the background model are
\begin{equation}
\bar\rho=1,\quad \bar c_s=1,\quad \bar B = 0.5,\quad c_A =0.5,\quad
\theta=\pi/4.
\end{equation}
The amplitude of the Alfven wave (in chosen dimensionless variables)
traveling in the positive direction of the $x$-axis is
$\hat{\mi{U}}_{+A}=(0,\:0,\:-1,\:0,\:1,\:0,\:0)^T$. For the slow and
fast MHD waves traveling in the same direction these amplitudes are
$\hat{\mi{U}}_{+S}=(1,\:0.33108,\:0,\:2.6894,\:0,\:-2.5184,\:1)^T$
and
$\hat{\mi{U}}_{+F}=(1,\:1.0679,\:0,\:-0.13146,\:0,\:0.39708,\:1)^T$
respectively. The cyclic frequencies of these waves are
$\omega_A=kV_A$, $\omega_S=kV_S$, and $\omega_F=kV_F$ respectively.
The results of our numerical and analytical solutions for the moment
of time $t$ = 20 min. for all three waves are shown in Figure
\ref{Fig:1Dtest_PlnWaves}. Panels a, b, and c represent the results
for the Alfven, slow, and fast MHD waves respectively. Only
variables $v$ and $B_y$ are nonzero in the Alfven wave. They are
shown by solid and dashed curves in panel a) respectively. The exact
solution for $v$ is shown by circles. In the slow and fast MHD waves
variables $\rho$, $u$, $w$, $B_z$, and $p$ are all nonzero. The
dimensionless pressure coincides in amplitude and phase with the
density and is not shown in Figure \ref{Fig:1Dtest_PlnWaves}.
Variables $\rho$, $u$, $w$, and $B_z$ are shown in panels b) and c)
by solid, dash-dotted, dashed, and dotted curves respectively. The
exact analytical solution for $w$ is shown by circles. The
analytical solutions for other variables coincide with the numerical
curves and thus are not shown. We see that our numerical solutions
reproduce the amplitudes, phases, and velocities of the Alfven,
slow, and fast MHD waves very well.

\section{Results and discussion}
\subsection{MHD waves generated by a single source in uniform and non-uniform
background magnetic field} 
\begin{figure}
\epsscale{1.0}\plotone{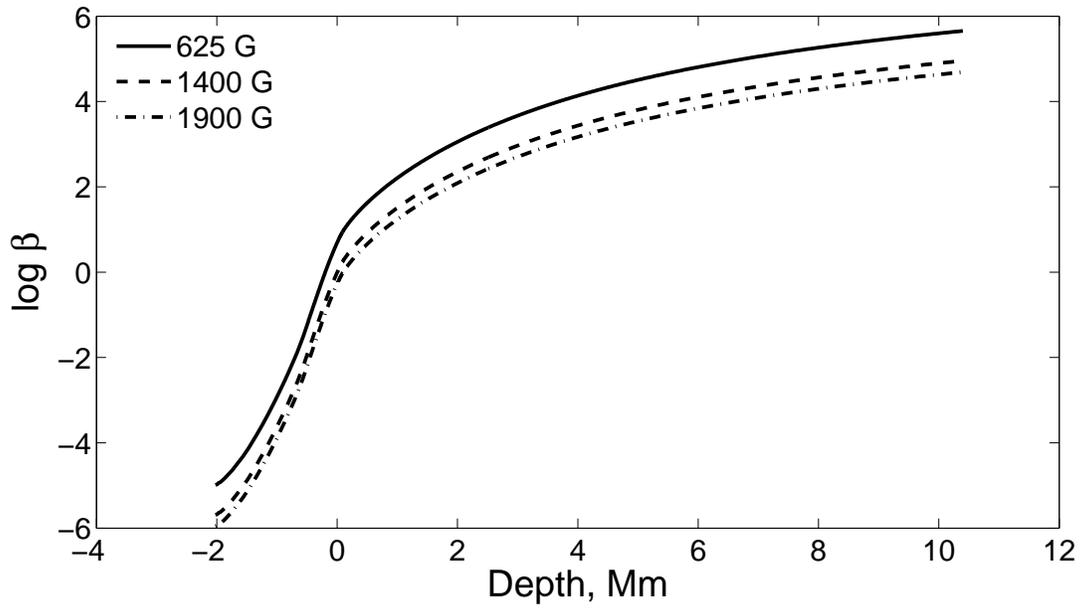} \caption{\label{Fig:beta} The ratio
of the gas pressure to the magnetic pressure (plasma parameter
$\beta$) as a function of depth for the background model for three
values of the magnetic field strength.}
\end{figure}

In this section we present our results of
numerical simulation of excitation and propagation of MHD waves
generated by a single source of vertical force with central
frequency $\nu$ = 3.5 mHz placed at depth $h_{src}=100$ km in a
rectangular region of size 15.5$\times$15.5$\times$12.5 Mm$^3$
(104$\times$104$\times$70 nodes). The horizontal grid is uniform
with $\Delta x=\Delta y = 150$ km. The vertical grid is non-uniform.
The grid step $\Delta z$ varies from 50 km near the photosphere to
600 km near the bottom of the computational domain. Time step
$\Delta t=0.5$ s was chosen to satisfy the Courant stability
condition. Vector $\mi{B}_0=(B_0 \sin\gamma,0,B_0\cos\gamma)^T$ of
the uniform inclined background magnetic field lies in XZ plane and
has inclination angle of $\gamma$ = 45$^\circ$ with respect to the
top boundary normal. The magnetic field strength, $B_0$, varied from
0 to 2500 G in our simulations. The ratio of the gas pressure to the
magnetic pressure, plasma parameter $\beta$ for typical sunspot
penumbra values is shown in Fig.~\ref{Fig:beta}.

We describe here two examples of wave propagation in the uniform
background magnetic field with a wave source located inside the
magnetic region, and of a non-uniform magnetic field with the
source located outside the magnetic regions. Since the convective
motions on the Sun are inhibited in strong magnetic field regions,
the acoustic sources there are suppressed. Thus, the second case
better describes the realistic situation on the Sun.

\begin{figure}
\epsscale{0.8}\plotone{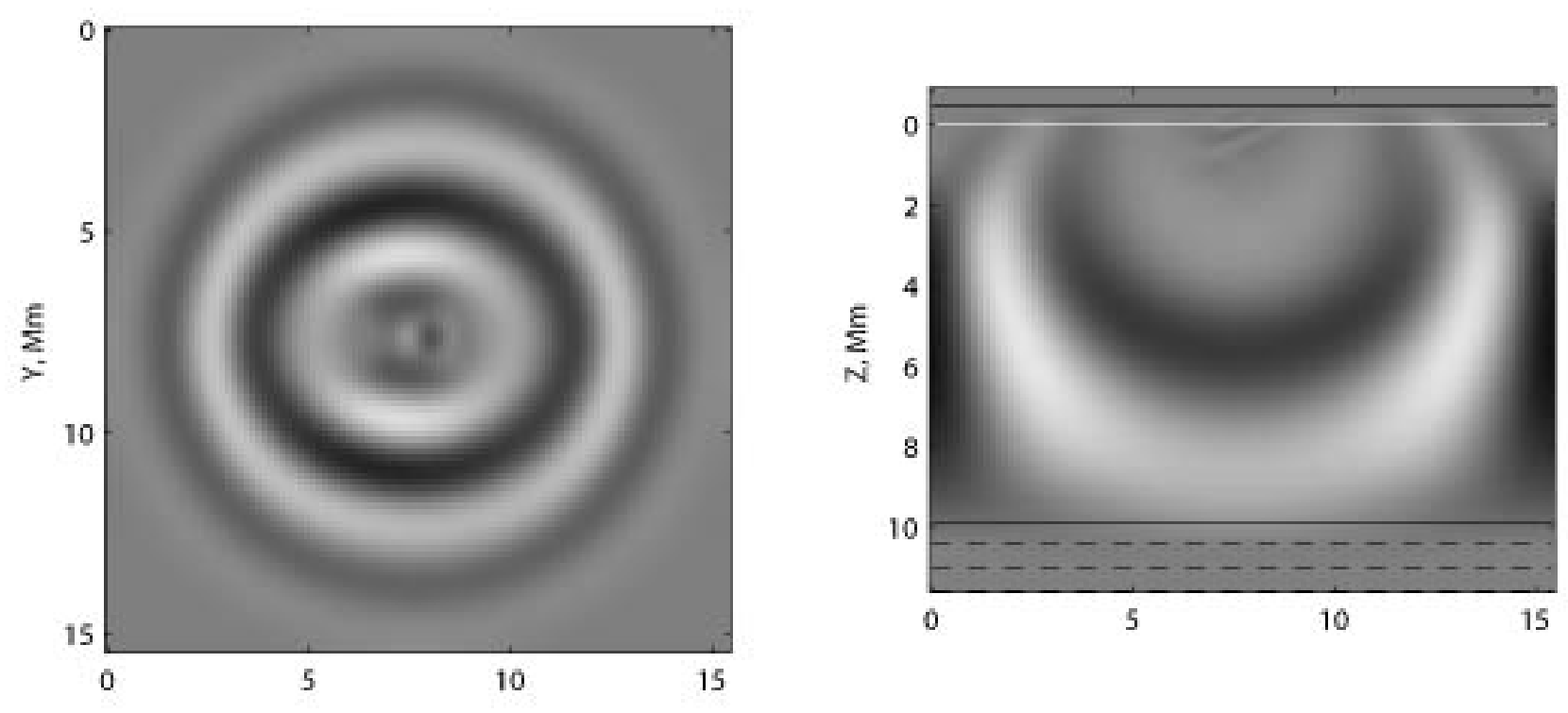}
\epsscale{0.8}\plotone{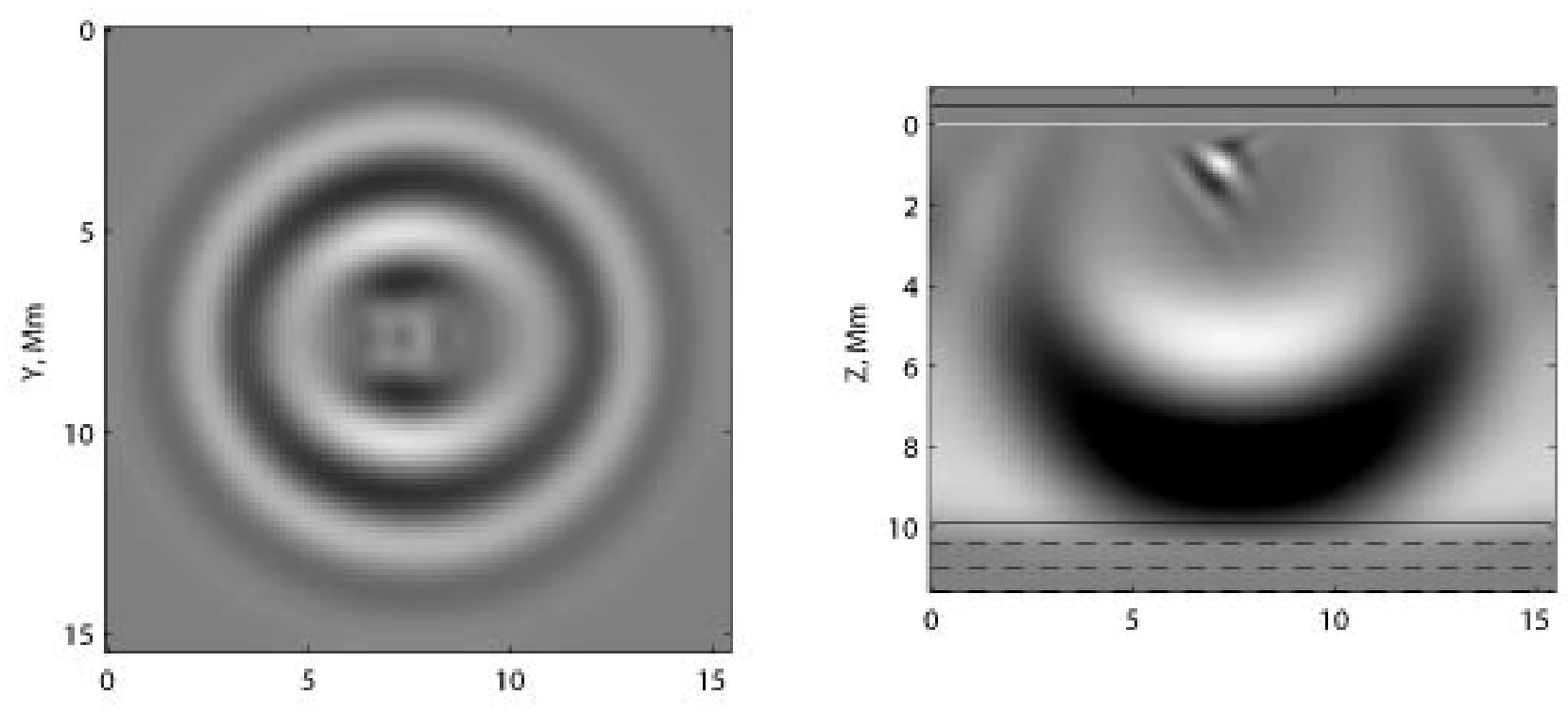}
\epsscale{0.8}\plotone{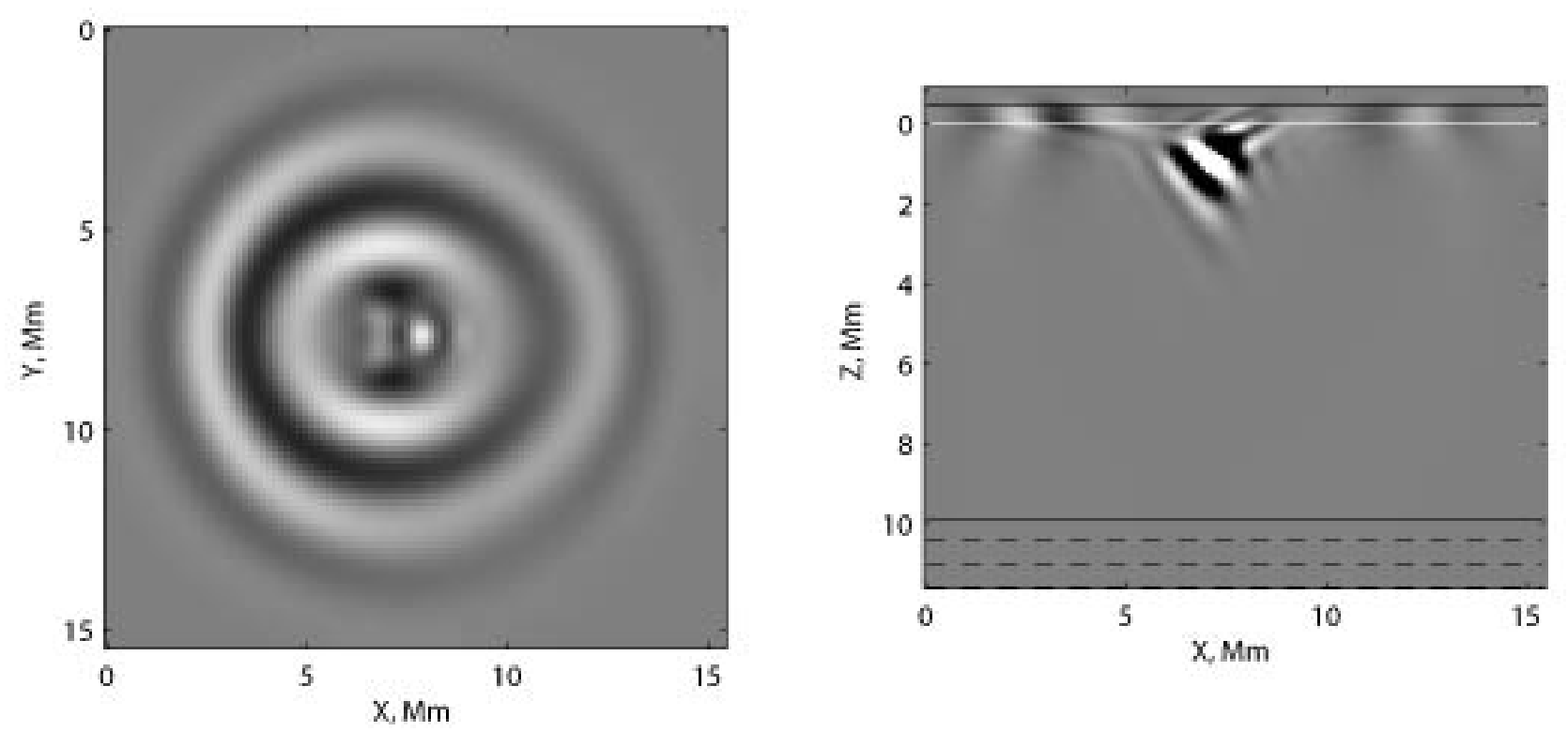}
\caption{\label{Fig:SingleSrc_UniB}
Snapshots of horizontal (left column) and vertical (right column)
slices of the 3D domain for perturbations of density $\rho$ (a),
z-momentum $\rho_0 w$ (b), and z-component of the magnetic field
$B_z$ (c). Strong Alfven wave is generated due to the interaction of
the wave source and the background magnetic field at the source
location. In this example, $B_0=625$ G, and inclined by $45^\circ$
in the XY-plane.}
\end{figure}

The simulation results for the uniform inclined magnetic field of
$B_0 = 625$ G are shown in Fig.~\ref{Fig:SingleSrc_UniB}. The
vertical map of $B_z'$ (panel c, right) reveals a strong Alfven
wave, which propagates along the background inclined magnetic field
lines. As expected, the Alfven wave is not presented in the map of
density perturbations. The Alfven wave is generated due to the
interaction between the wave source and the magnetic field at the
source location. The concentric waves in the left panels represent a
mixture of the fast MHD wave (analogous to p-modes in absence of the
magnetic field) and the surface magnetic-gravity wave (analogous to
f-mode). Since the wave speed depends on the angle between the
vectors of magnetic field vector and wavenumber, the wave fronts are
anisotropic. The separation of these two types of waves will be
discussed in Sec.~3.2 together with analysis of phase and group
travel times.

\begin{figure}
\epsscale{0.8}\plotone{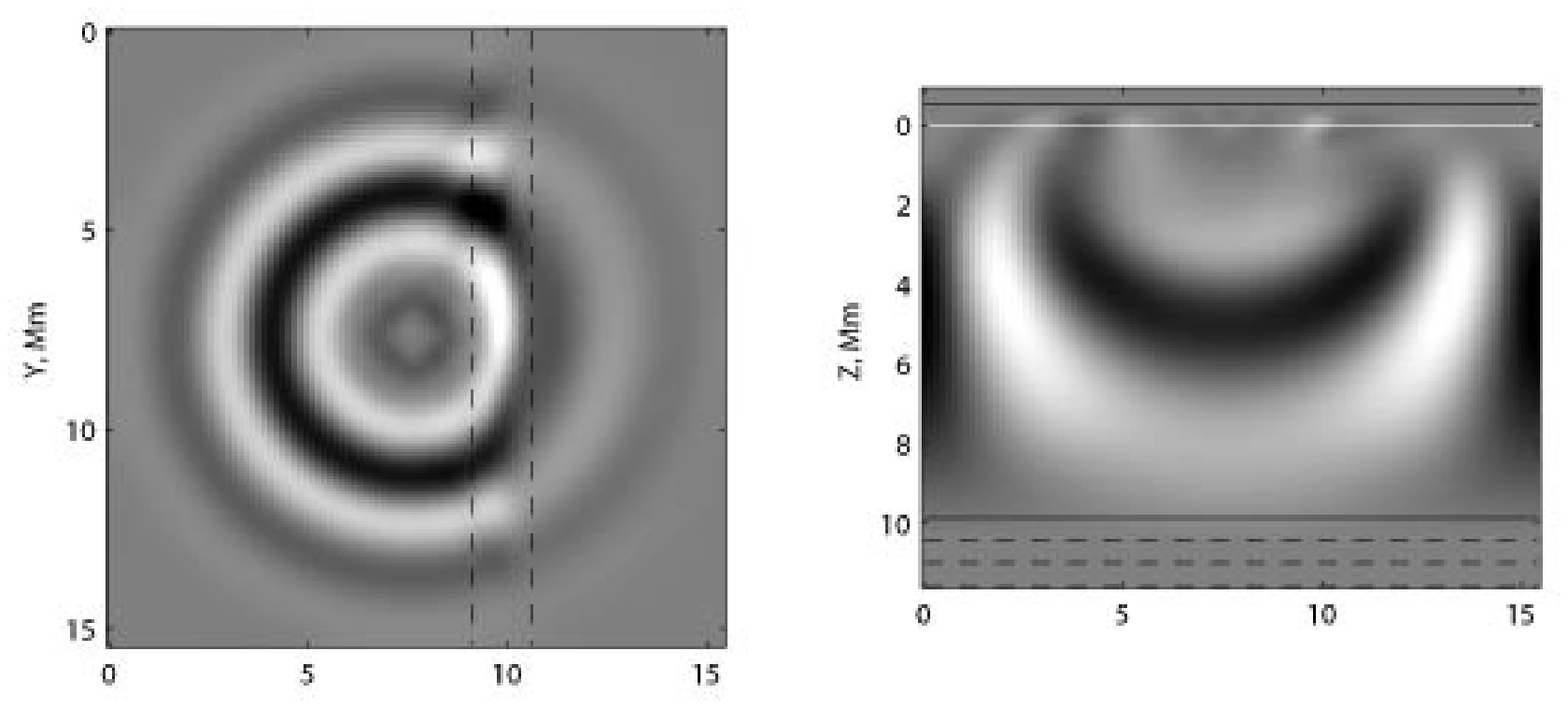}
\epsscale{0.8}\plotone{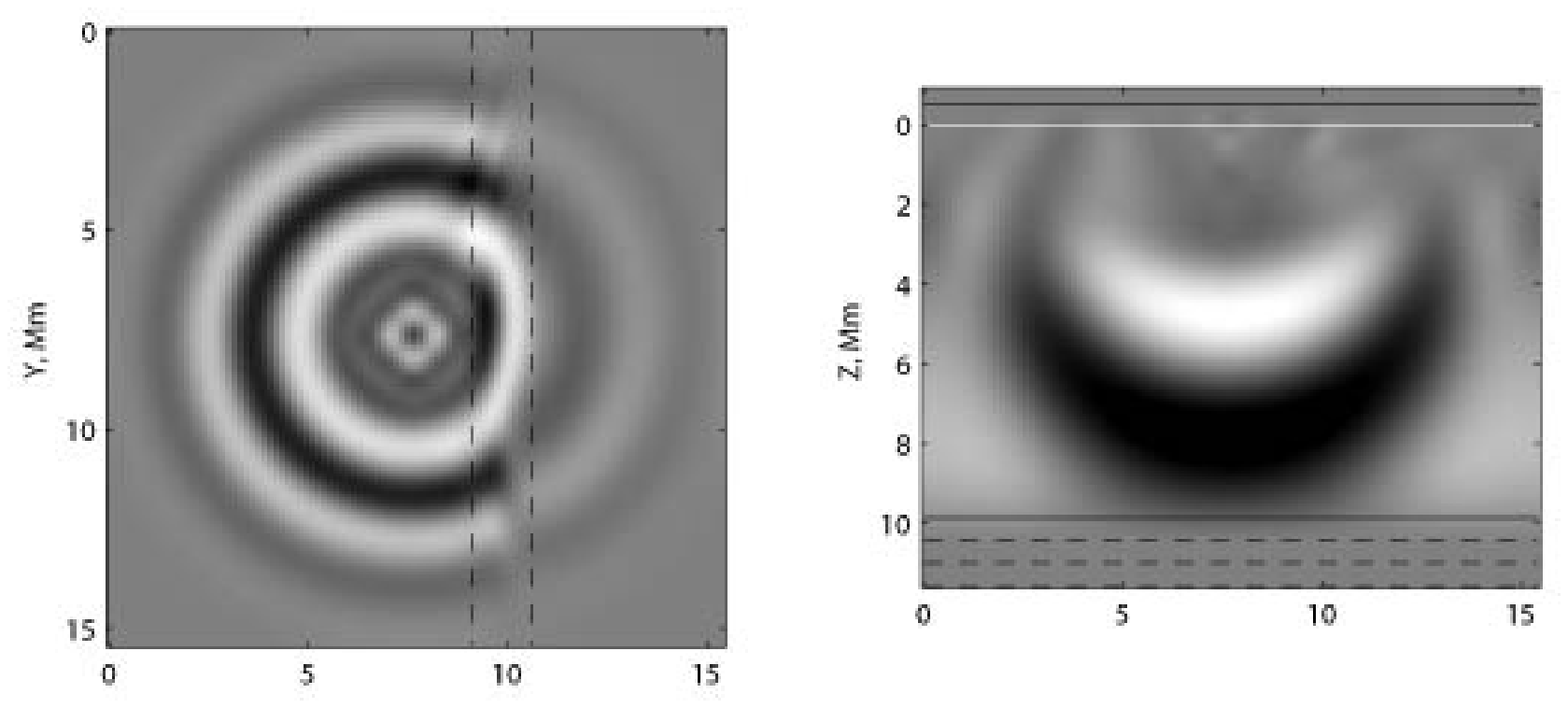}
\epsscale{0.8}\plotone{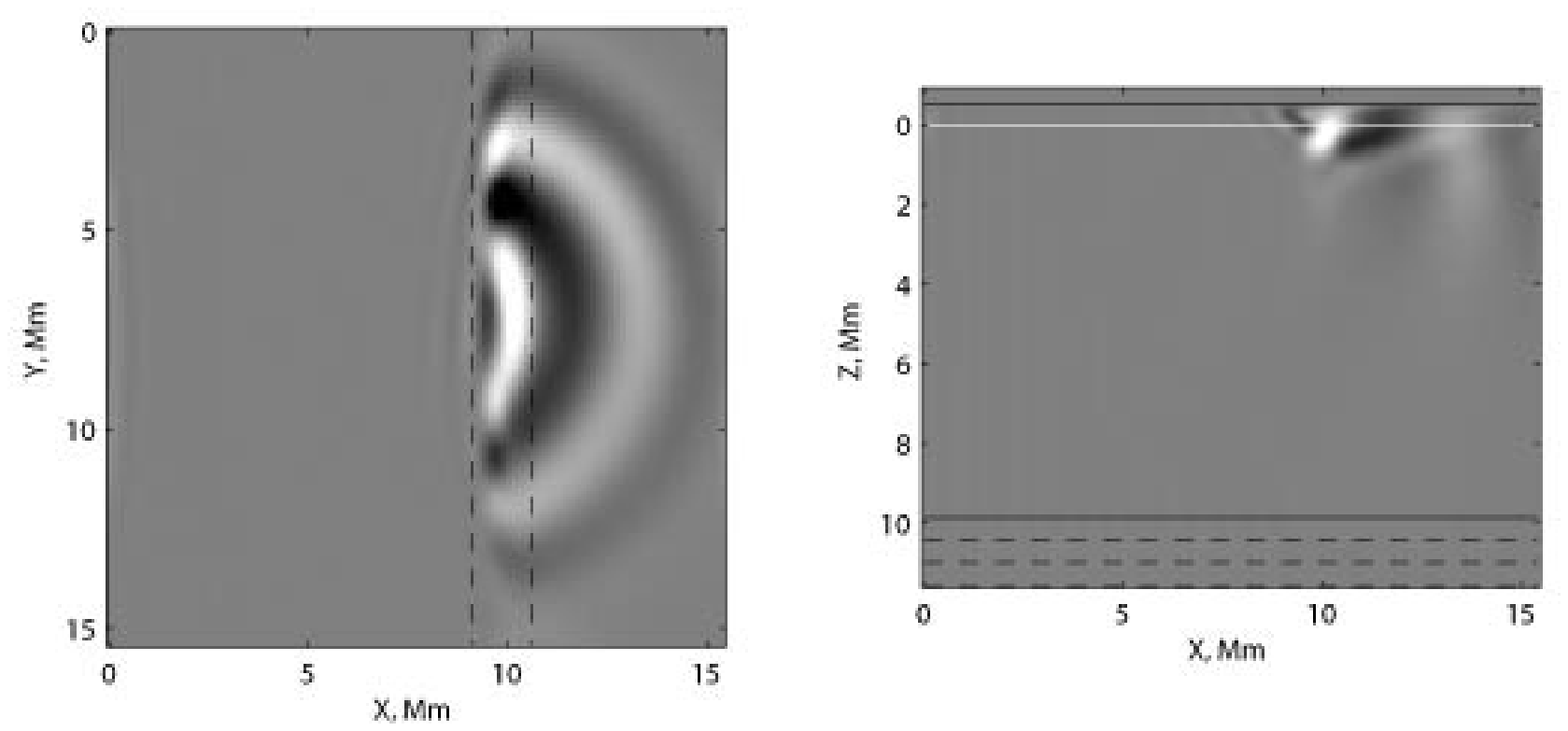}
\caption{\label{Fig:SingleSrc_NonUnB} Snapshots of horizontal (left
column) and vertical (right column) slices of the 3D domain for
perturbations of density $\rho$ (a), z-momentum $\rho_0 w$ (b), and
z-component of the magnetic field $B_z$ (c). The strength of the
background magnetic field (inclined by $45^\circ$ in the YZ-plane)
is 2500 G in the region to the right from vertical dashed lines in
left column and 0 in the left region. The wave source is located in
the region free of magnetic field. Propagating into the region with
the background magnetic field acoustic and surface gravity waves are
transformed into the fast MHD wave and magneto-gravity wave. The
Alfven wave does not appear in this case.}
\end{figure}

The simulation results for the second case, when the source is
placed outside the region containing the background magnetic field
are shown in Figure \ref{Fig:SingleSrc_NonUnB}. The domain size and
grid parameters are the same, as in Figure \ref{Fig:SingleSrc_UniB}.
The background magnetic field $\mi{B}_0=(0,B_0
\sin\gamma,B_0\cos\gamma)^T$ lies in YZ plane, has inclination angle
$\gamma$ = 45$^\circ$ with the top boundary normal and is parallel
to the planes showed in left columns of figure  by dashed lines. The
magnetic field $B_0$ has constant values of 0 and 2500 G in the
regions on the left and right of the dashed lines respectively.
Between the dashed lines the strength of the background magnetic
field is linearly decreases from maximum value to zero. Such
configuration of the background field satisfies condition
$\mbox{div} \mi{B}_0=0$. The magnetic field strength in this example
is chosen higher than the typical penumbra value for a better
demonstration of the magnetic effects.
 The wave source is placed in the region
free from the background magnetic field. The waves generated by such
a source are pure acoustic and surface gravity waves. When they
enter the region occupied by the inclined magnetic field they are
transformed into the fast MHD and slow magneto-gravity waves
respectively. The Alfven wave does not appear in these simulations.
Evidently, the fast MHD wave travels faster than the original
acoustic wave, and its amplitude is reduced. In these simulations we
do not notice significant wave transformation effects in the near
surface reflection layers where the plasma parameter, $\beta$, equal
to 1. The helioseismic waves are trapped below the surface and
according to our simulations are not affected by transformation into
other types (slow MHD and Alfven modes). We will discuss in more
detail the role of the transformation for waves of different
frequencies in a future publication. The main purpose of this paper
is to discuss the effects of the inclined magnetic field on the
observed travel time variations in the sunspot penumbra.

\subsection{Phase and group travel time variations along the wavefront}
\begin{figure}
\epsscale{1.0}\plotone{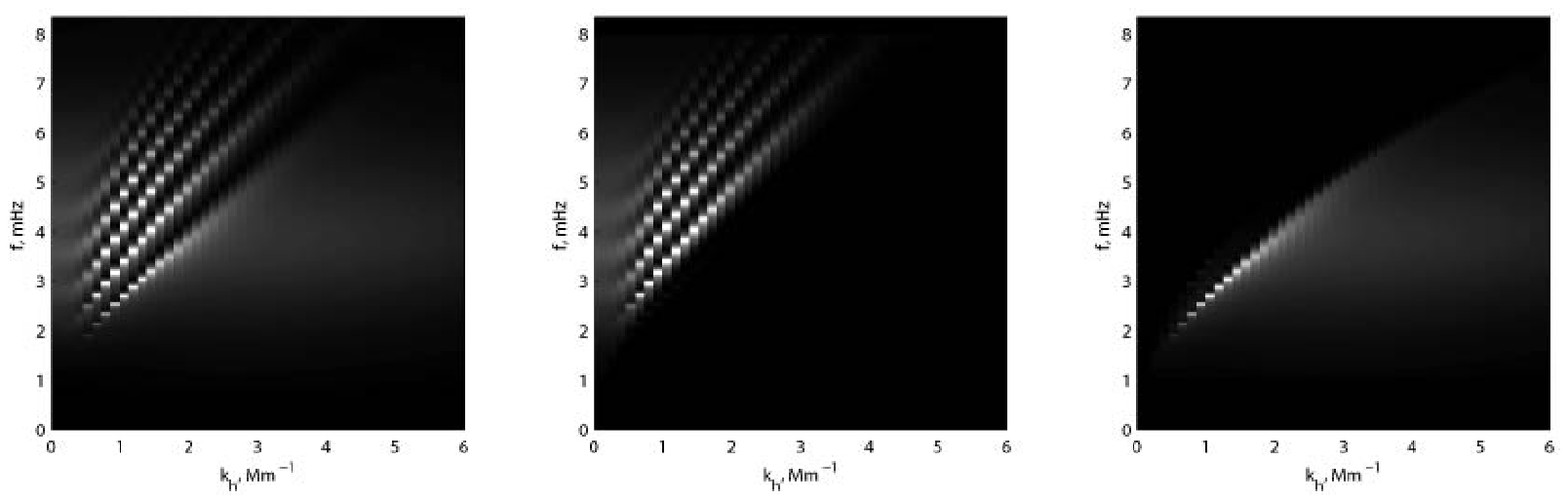}
\epsscale{1.0}\plotone{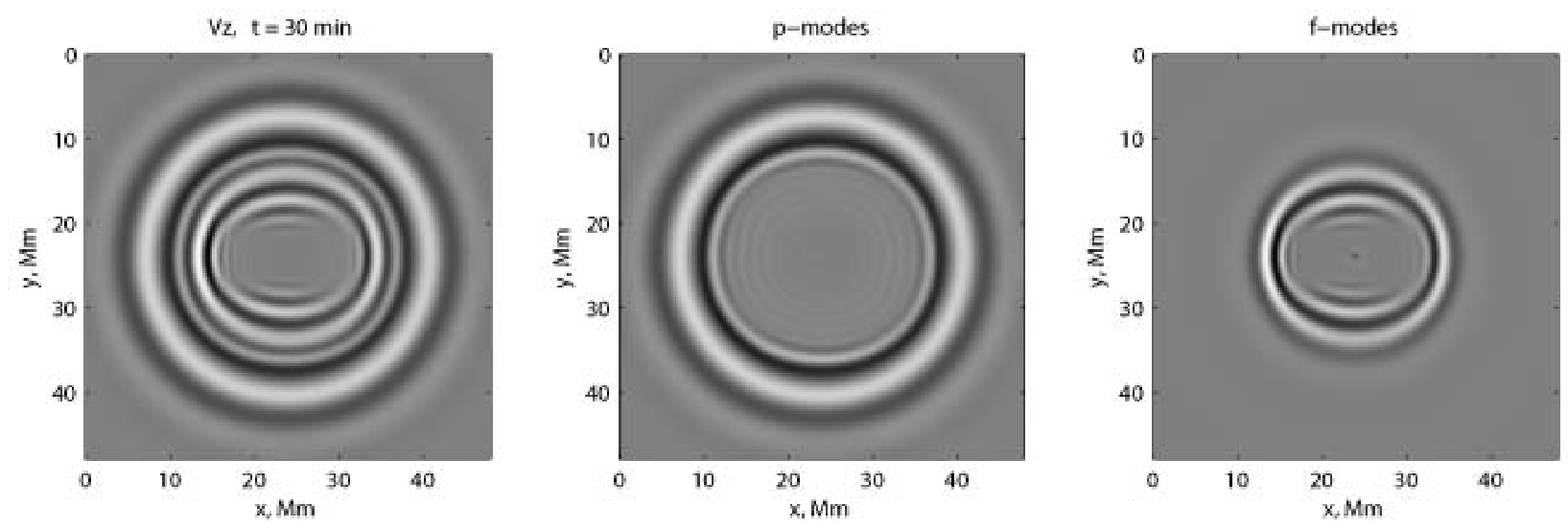}
\caption{\label{Fig:kw_diagr} Spectra ($k$--$\nu$ diagrams) and
corresponding maps of perturbation of the z-component of velocity
are shown on top and bottom rows respectively. Results for the
original MHD wave field and wave fields after filtering out $f$- and
$p$-modes are shown on panels a, b, and c respectively.}
\end{figure}

\begin{figure}
\epsscale{1.0}\plotone{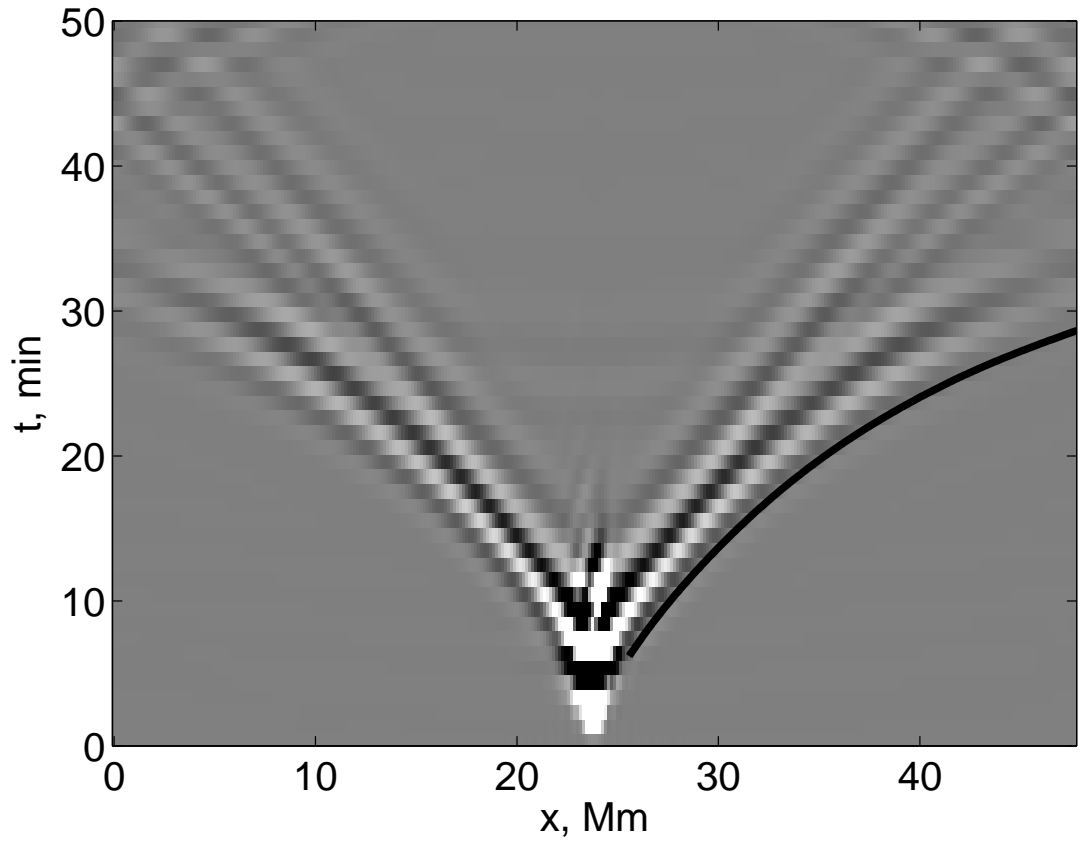}
\caption{\label{Fig:rho_TD}
Time-distance diagram for density perturbations. Solid curve
represents a theoretical time-distance curve for $p$-modes and the
standard solar model in absence of the magnetic field. Fast MHD wave
and magnetic-gravity waves are separated due to the different
dispersion relations.}
\end{figure}

To study the travel time variations we performed simulations in a
rectangular box of size 48$\times$48$\times$12.5 Mm$^3$
(320$\times$320$\times$70 nodes) for different values $B_0$ and
inclination angles $\gamma$ of the uniform background magnetic
field: (625 G, 70$^\circ$), (1400 G, 45$^\circ$), and (1900 G,
30$^\circ$). The grid step size, source type and depth were chosen
the same as in previous Section 3.1. Total simulation time equals 6
hours of solar time. The simulation results for $B_0$=625 G and
$\gamma$=70$^\circ$ are shown in Figure \ref{Fig:kw_diagr}. The top
row represents $k$--$\nu$ diagrams, the bottom one shows
corresponded horizontal snapshots of the z-component of velocity at
the level of the photosphere for the moment of time $t=30$ min. The
usual technique of fitting of the cross-covariance function by
Gabor's wavelet was used to calculate the travel times. This
technique was developed for $p$-modes \citep{Kosovichev1997}. The
source of the vertical (z-component) of force generates a strong
gravity wave (the lowest ridge of $k$--$\nu$ diagram on panel a
corresponds to the $f$-mode) which has to be filtered out. Results
of separation of $p$- and $f$-modes are shown in panels (b) and (c)
respectively. The maps of $V_z$ for $p$- and $f$-modes (bottom row)
were obtained by taking the inverse Fourier transform of the
corresponding 3D spectra. It is clear, even without applying an
$f$-mode filter, that starting form the distance of about 15 Mm from
the source the $p$- and $f$-modes are spatially separated due to the
different velocities. Due to the different dispersion relations the
magnetic-gravity and fast MHD waves easily separated on
time-distance diagram (see Figure \ref{Fig:rho_TD}). The solid black
curve represents a theoretical time-distance curve for $p$-modes and
standard solar model in absence of the magnetic field. We see, that
the fast MHD waves are not significantly affected by the magnetic
field and follow the theoretical $p$-mode curve, calculated for the
quiet Sun in the ray approximation \citep{Kosovichev1997}. The
magnetic-gravity waves have characteristic "zebra" structure due to
the difference between phase and group velocities. Their speed is
less than the speed of the fast MHD wave.

\begin{figure}
\epsscale{1.0}\plotone{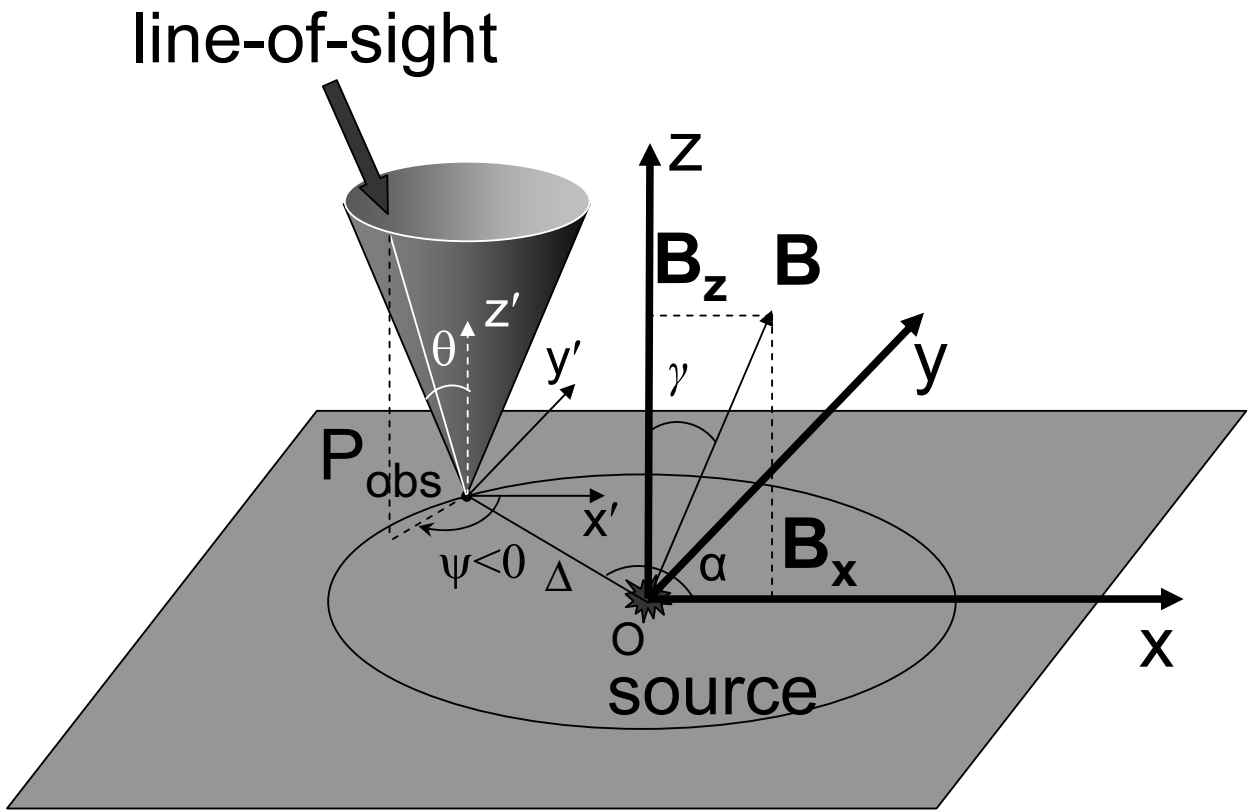}
\caption{\label{Fig:LoS_geometry}
Choice of the coordinate system. The line-of-sight direction is
defined by angles $\psi$ and $\theta$. Position of observational
point $P_{\mathrm{obs}}$ on the photosphere is fixed by azimuthal
angle $\alpha$ and distance $\Delta$ from the projection of wave
source O.}
\end{figure}

The magnetic field results in anisotropy of the wave properties.
Therefore, the wave travel times measured from the line-of-sight
component of the displacement velocity depend on the direction of
the wave propagation and also on the viewing angle. We have
investigate this effect for the case of the uniform inclined
magnetic field. The choice of the coordinate system and geometry are
shown in Figure \ref{Fig:LoS_geometry}. Horizontal XY-plane
coincides with the photosphere. The origin of the Cartesian
coordinate system is placed at the point O above the wave source
(the source itself is at the depth of 100 km below the photosphere).
The uniform inclined background magnetic filed lies in the XZ-plane
($B_y=0$) and has angle $\gamma=45^\circ$ with the normal to the
photosphere. The location of point of observations
$P_{\mathrm{obs}}$ is defined by the distance $\Delta$ from the wave
source and the azimuthal angle $\alpha$. The line-of-sight (LoS)
direction is defined by two angles: angle $\theta$ between LoS
direction and local normal, and azimuthal angle $\psi$ between OX
axis and projection of the LoS on the local horizontal plane. First,
we build 3D $k$--$\nu$ diagrams for each velocity component $u$,
$v$, and $w$, filter out the $f$-modes as shown in Figure
\ref{Fig:kw_diagr} and calculate the Cartesian components of
velocity perturbation $u_p$, $v_p$, and $w_p$ for $p$-modes by
taking the inverse Fourier transform of corresponding spectra. Then,
we calculate cross covariance
\begin{equation}
C(\mi{r}_1,\mi{r}_2,\tau)=\frac{1}{T}\int_0^T
v_{LoS}^{(p)}(\mi{r}_1,t) v_{LoS}^{(p)}(\mi{r}_2,t+\tau) dt
\end{equation}
of LoS velocities
\begin{equation}
v_{LoS}^{(p)}=u_p\cos\psi\sin\theta + v_p\sin\psi\sin\theta +
w_p\cos\theta
\end{equation}
in the origin of the system of coordinates and in the observational
point $P_{\mathrm{obs}}$ ($\Delta=7.9$ Mm). Then we fit the cross
covariance function with Gabor's wavelet
\begin{equation}
G(\Delta,\tau)=A\cos[\omega_0(\tau-\tau_p)]\exp\left[
-\frac{\delta\omega^2}{4}(\tau-\tau_g)^2\right],
\end{equation}
where $\Delta=|\mi{r}_1 - \mi{r}_2|$ is the distance between points
where LoS velocities $v_{LoS}^{(p)}(\mi{r}_1,t)$ and
$v_{LoS}^{(p)}(\mi{r}_2,t)$ are measured, $A$ is the amplitude,
$\omega_0$ is the central frequency, $\tau_p$ and $\tau_g$ are the
phase and group travel times respectively, and $\delta\omega$ is the
bandwidth. Parameters $A$, $\omega_0$, $\tau_p$, $\tau_g$, and
$\delta\omega$ are free and have to be determined from the fitting
procedure. Repeating this procedure for different observational
points with the same $\Delta$ but different azimuthal angle we
obtain $\tau_p$ and $\tau_g$ as functions of $\alpha$.

\begin{figure}
\epsscale{0.8}\plotone{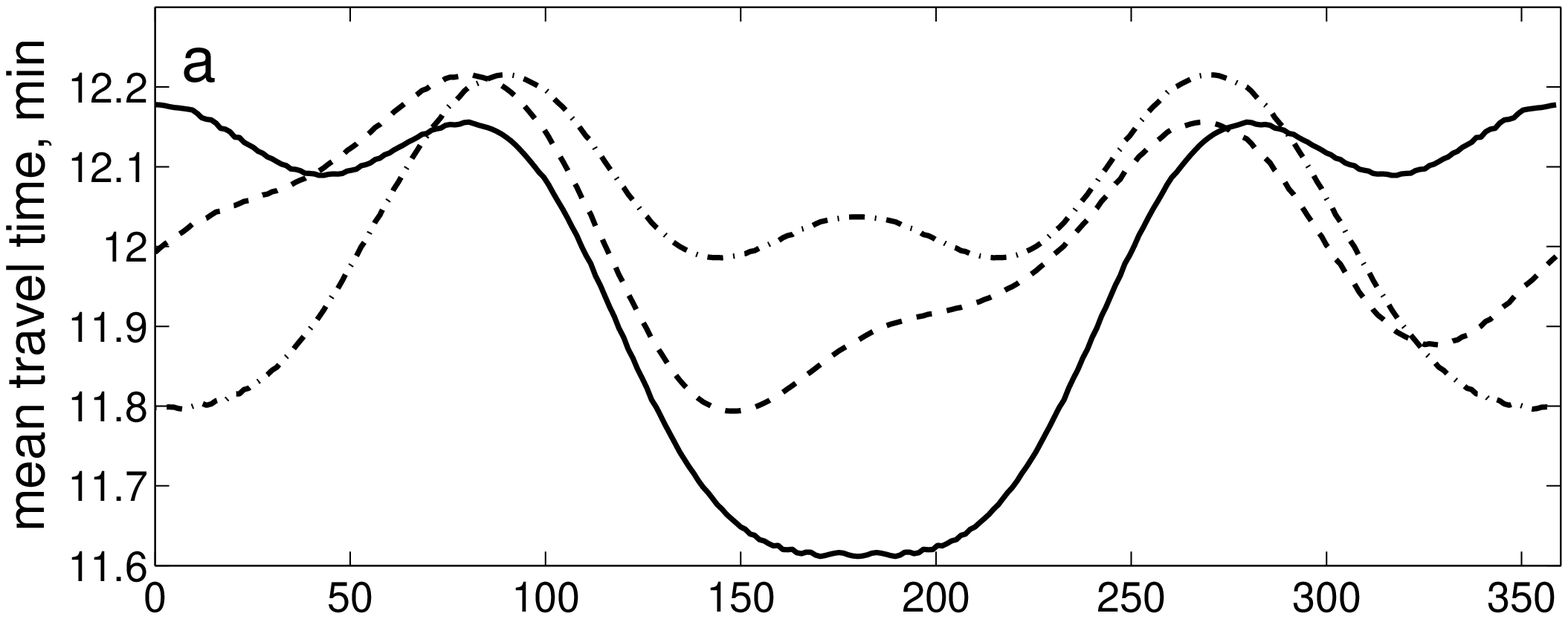}
\epsscale{0.8}\plotone{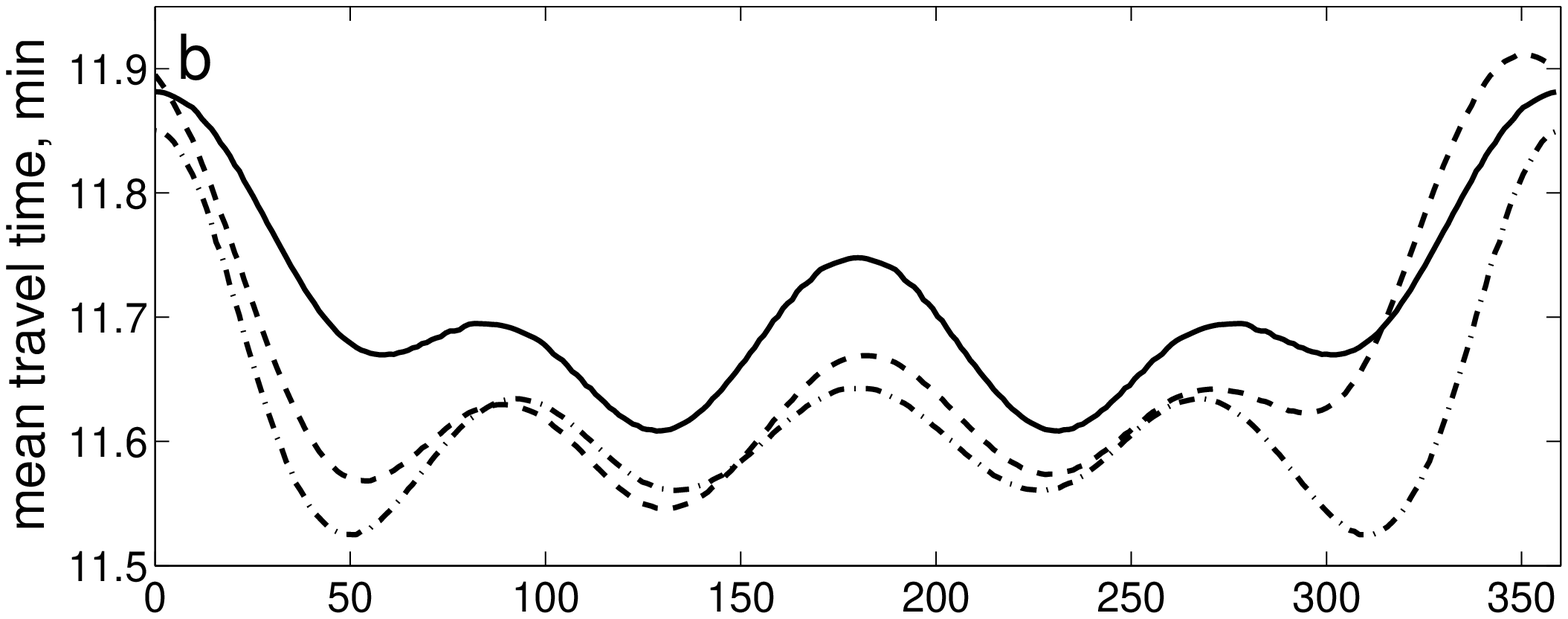}
\epsscale{0.8}\plotone{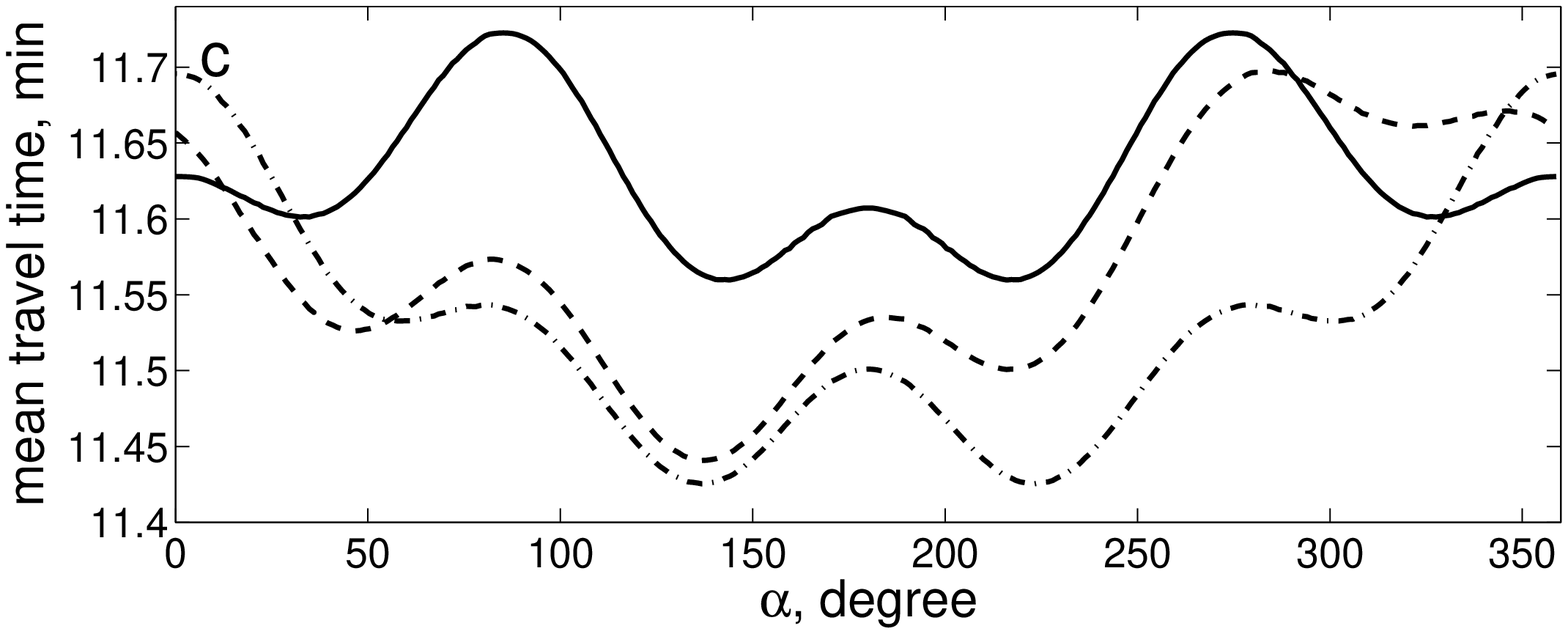}
\caption{\label{Fig:TravelTimes_alph} Mean phase travel times vs.
azimuthal angle $\alpha$ calculated from Doppler LoS velocities at
the hight of 300 km above the photosphere for different strengths of
the uniform inclined background magnetic field (625 G for panel a,
1400 G for panel b, and 1900 G for panel c), and different LoS
directions. Solid, dashed, and dash-dotted curves represent travel
time variations for $\psi$ = $0^{\circ}$, $90^{\circ}$, and
$180^{\circ}$ respectively.}
\end{figure}

Travel time variations along the wave front for different $B_0$ and
different LoS angles are shown in Figure \ref{Fig:TravelTimes_alph}.
Panels a, b, and c correspond to $B_0$ of 625 G, 1400 G, and 1900 G
respectively. The solid, dashed, and dash-dotted curves correspond
to $\psi$ = \{0$^\circ$, 90$^\circ$, 180$^\circ$\} respectively. The
angle between the LoS and the local normal is $\theta=20.5^{\circ}$.
The mean phase travel times show variations of about 0.5 min along
the wave front, the variation amplitude is a little smaller for
strong (1400--1900 G) magnetic fields than for the weak 625 G field.
For the weak magnetic field changes of $\psi$ change the shape of
the curve, but not its average value. The strong magnetic fields
cause anisotropy, and the mean value of the travel times changes
with angle $\psi$. Thus, averaging along the wave front for the
strong magnetic fields gives variations of the observed mean travel
times of about 0.1--0.3 min with angle $\psi$ (see detailed
discussion in the next section).

\subsection{Azimuthal dependence of phase travel times in sunspots:
comparison with observations} \citet{Zhao2006} observed different
behavior of azimuthal dependence of phase travel times obtained from
the SOHO/MDI Doppler shift data \citep{Scherrer1995} in sunspots
depending on their position on the disk. As far as we want to
reproduce similar conditions (and geometry) in our simulations we
give a detailed description of their algorithm of the phase travel
time calculation. We will apply the same technique to our simulated
data.

\begin{figure}
\epsscale{1.0}\plotone{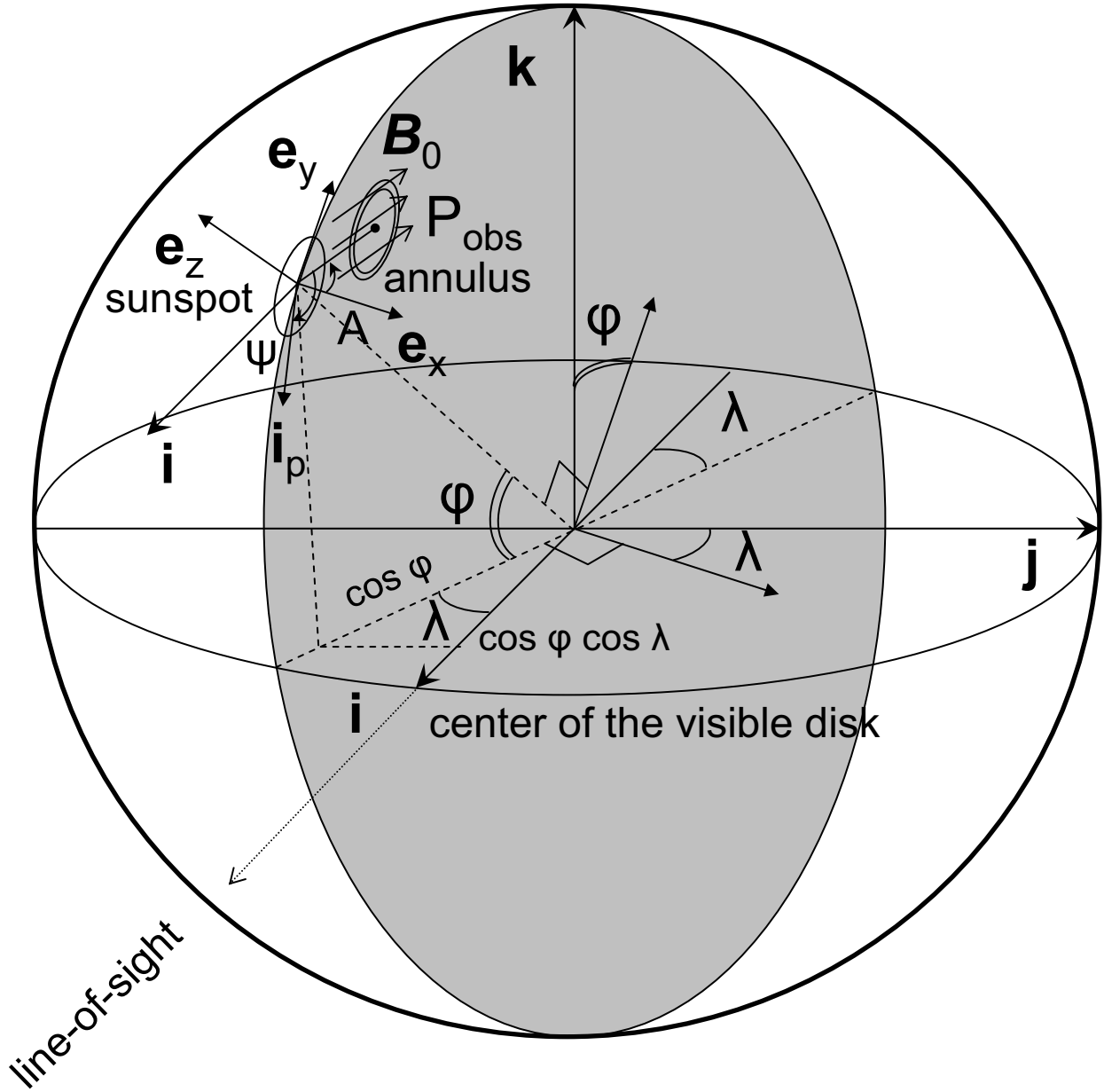}
\caption{\label{Fig:SunSpot_geometry}
To derivation of the relation between $\psi$ and $A$. Angle $A$ is
the azimuthal angle of the observation point in the local coordinate
system associated with the sunspot. Angle $\psi$ is one of two
angles determining the LoS direction.}
\end{figure}

The sunspot is located at latitude $\phi$ and longitude $\lambda$ on
the eastern part of the disk as shown in Figure
\ref{Fig:SunSpot_geometry}. The sunspot meridian plane is shown by
gray color. Orth $\mi{i}$ of the global system of coordinate with
the origin in the center of the Sun is aimed at the center of the
visible solar disk, orth $\mi{k}$ points to the north, and orth
$\mi{j}$ points to the west. The local system of coordinates with
the origin in the center of the sunspot is chosen in such a way that
$\mi{e}_z$ coincides with the local normal to the surface,
$\mi{e}_x$ is directed to the west, and $\mi{e}_y$ is directed along
the meridian and forms a right vector triplet with $\mi{e}_z$ and
$\mi{e}_y$.

Observational point $P_{\mathrm{obs}}$ is chosen inside the sunspot
penumbra. Azimuthal angle $A$ is counted counterclockwise from local
west direction $\mi{e}_x$. Two signals are calculated: the LoS
velocity at $P_{\mathrm{obs}}$ and the LoS velocity averaged along
the annulus (between inner and outer radii of a ring with average
radius $\Delta$ = 8 Mm with the origin in the observational point).
Cross covariance of these two signals is fit with the Gabor's
wavelet. Fitting procedure gives us mean phase and group travel
times. We assume, that the annulus is small enough that sunspot
magnetic field $\mi{B}_0$ inside the annulus can be considered as
uniform. Hence the problem is reduced to simulations with the
uniform inclined magnetic field shown in Figure \ref{Fig:kw_diagr}.
To compare results of the simulations with the observations we have
to find from what direction we have to look at our simulation domain
to match the observations. In other words, we have to find a
relation between azimuthal angle $A$ of the sunspot center and
$\psi$. Angle $\psi$ here has the same meaning as in Figure
\ref{Fig:LoS_geometry}. This is the angle between the projection of
the background magnetic field on the local horizon (XY-plane) and
projection of the LoS on the same plane. The unit vector in LoS
direction in the sunspot local system of coordinates is given by
equation
\begin{equation}
\mi{i}=\sin\lambda\;\mi{e}_x -\sin\phi\cos\lambda\; \mi{e}_y +
\cos\phi\cos\lambda\; \mi{e}_z.
\end{equation}
The projection of $\mi{i}$ on local horizontal plane (defined by
vectors $\mi{e}_x$ and $\mi{e}_y$) has coordinates
$\mi{i}_p=(-\sin\lambda,-\sin\phi\cos\lambda,0)^T$ (non-unit
vector). Angle $\psi$ is given by equation
\begin{equation}
\psi=-A-\angle\mi{i}_p\mi{e}_x=-A-\arccos\left(
\frac{-\sin\lambda}{\sqrt{\sin^2\lambda+\sin^2\phi\cos^2\lambda}}\right).
\end{equation}

\begin{figure}
\epsscale{0.8}\plotone{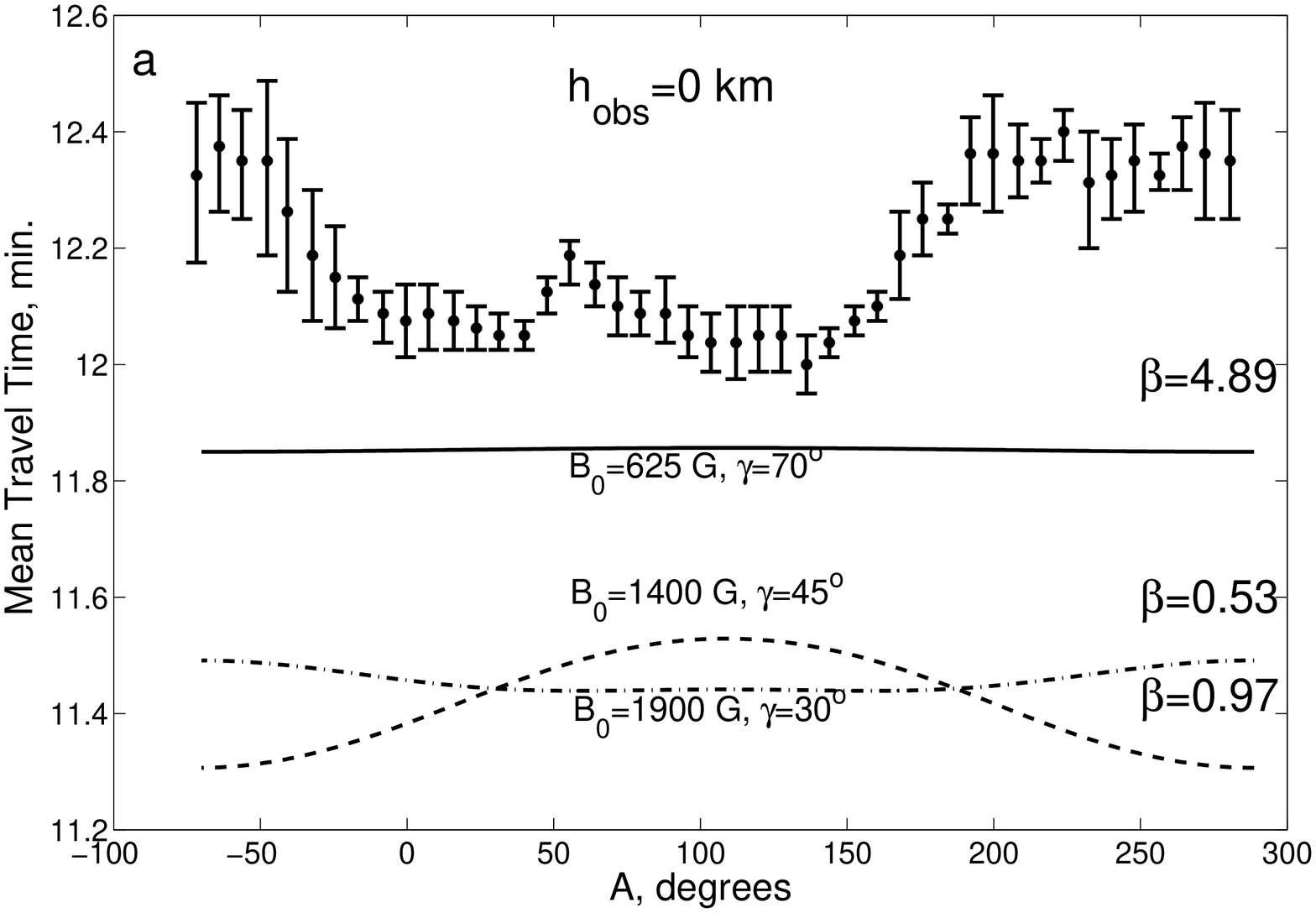}
\epsscale{0.8}\plotone{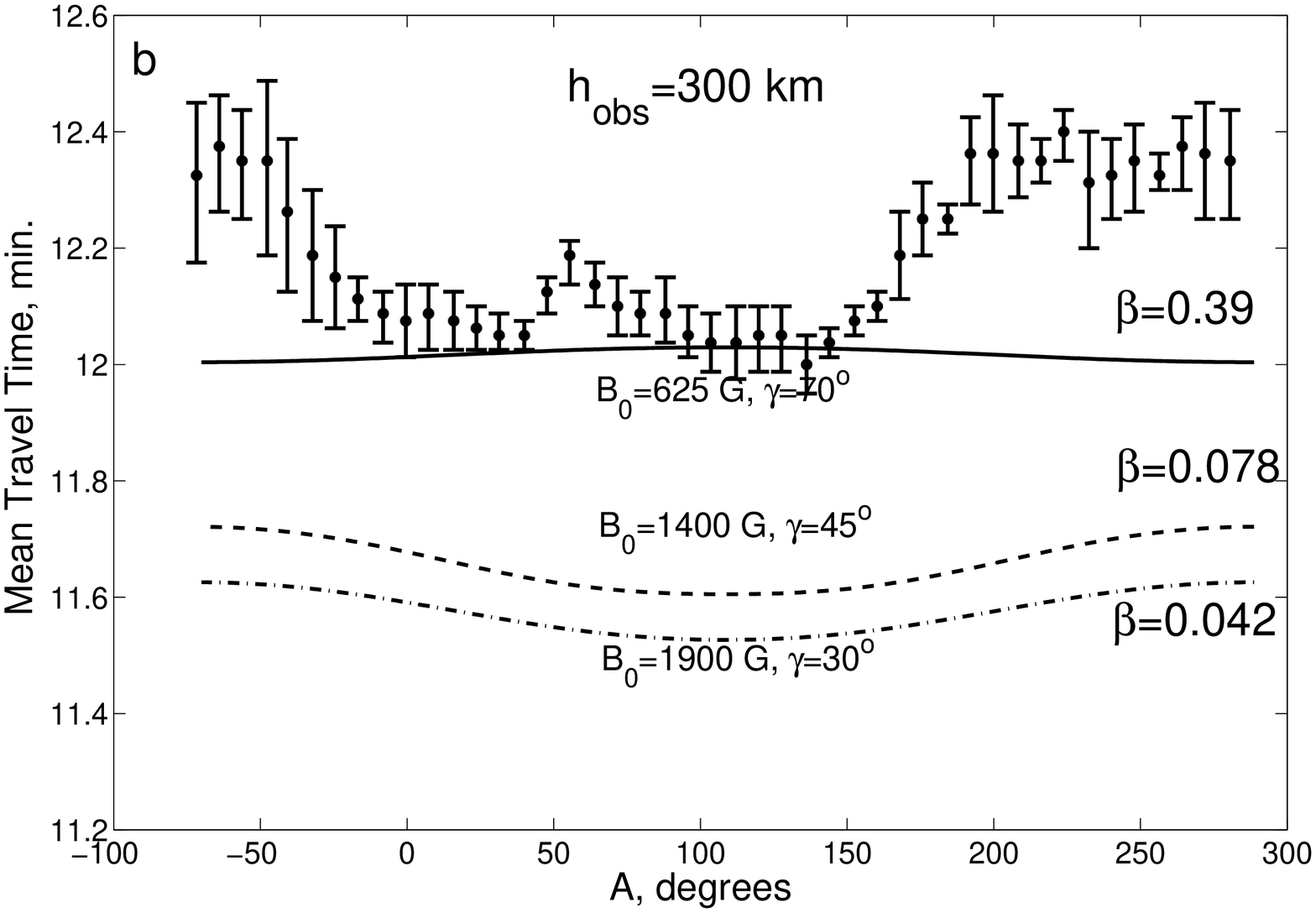}
\caption{\label{Fig:TravTimeAvr_A} Phase travel time obtained from
simulations averaged along with annulus with radius 8 Mm as a
function of azimuthal angle $A$.  The solid, dashed, and dot-dashed
curves corresponds to the following combinations of the magnetic
field strength and inclination angles: (i) $B_0=625$ G,
$\gamma=70^\circ$, (ii) $B_0=1400$ G, $\gamma=45^\circ$, (iii)
$B_0=1900$ G, $\gamma=30^\circ$. Data points with errorbars
represent observations.}
\end{figure}

We performed numerical simulations of propagation of MHD waves in
presence of the uniform inclined magnetic field for different
inclination angles and strengths of the magnetic field. The goal of
these simulations was to calculate contribution of the inclined
magnetic field effect to variations of the mean travel times with
the azimuthal angle. This is why we used horizontally uniform
standard solar model as a background model. The same model was used
to obtain Figures \ref{Fig:SingleSrc_UniB},
\ref{Fig:SingleSrc_NonUnB}. The mean travel times obtained for the
photospheric level (panel a) and the level of 300 km above the
photosphere (panel b) are shown in Figure \ref{Fig:TravTimeAvr_A}.
The solid curve corresponds to the magnetic field $B_0=625$ G with
inclination angle $\gamma=70^\circ$. The dashed and dash-dotted
curves are corresponded to cases $B_0=1400$ G, $\gamma=45^\circ$ and
$B_0=1900$ G, $\gamma=30^\circ$ respectively. The systematic shift
of the calculated mean travel times with respect to the observations
is caused by temperature effects (the sound speed is smaller inside
sunspots than in the quiet Sun), which are not included in these
simulations. The travel times systematically decrease when the
strength of the magnetic field increases because the speed of the
fast MHD wave increases with the magnetic field. For studying the
inclined field effect we are mostly interested in relative
variations of the mean travel times, provide the absolute values to
give a general impression about the magnitude of contributions to
the mean travel times from the temperature effect and variations of
the fast MHD wave speed with the magnetic field.  A model of a
sunspot with both temperature and magnetic field effect will be
presented in a future publication.

The amplitude of the mean travel time variations for weak (625 G)
magnetic field is 10 times smaller than the observed quantity for
the height of 300 km and even smaller for the photospheric level.
For stronger magnetic fields (1400--1900 G) the amplitude of travel
time variations is about 25\% of the observed amplitude. Behavior of
the travel times depends on the magnetic field strength and the
level of observation of Doppler velocities. For the weak magnetic
field (625 G) the phase of the travel time variations is opposite to
the observations for both levels. For the height of 300 km above the
photosphere where plasma parameter $\beta\ll1$ simulated travel
times show the same phase as in observations for both magnetic field
strengths (1400 G and 1900 G). For the photospheric level and field
strength of 1900 G the phase of travel time variations coincides
with the observations, while the travel time variations for
$B_0=1400$ G are in antiphase with the observations. Comparison of
the curves at different heights for the same inclination angles and
magnetic field strengths shows that for the photospheric layer the
mean travel times are systematically smaller by only about 0.12 min
than for the layer of 300 km above the photosphere. Thus, the
variations in the height of Doppler shift measurement do not have a
significant effect.

The curves in Figure \ref{Fig:TravTimeAvr_A} were obtained for the
annulus radius of 8 Mm and the sunspot located at heliospheric
latitude $\phi=19.5^\circ$ and longitude $\lambda=-6.5^\circ$. The
angle between LoS and local normal to the photosphere at the center
of the sunspot ($\theta$ angle) is $20.5^\circ$. The parameters are
chosen to match corresponded angles for Figure 1b of
\citet{Zhao2006}.

\section{Conclusion}
The numerical 3D simulations of excitation and propagation of
magneto-acoustic-gravity waves in the solar interior and atmosphere
show that the presence of the inclined background magnetic field
significantly alters properties of $f$-modes but has little effect
on $p$-modes. The interaction of the wave source with the background
magnetic field generates a strong Alfven wave. However, the Alfven
wave does not appear when the wave source is located outside the
magnetic region, and the acoustic-gravity waves propagate from the
region without magnetic field into the region with magnetic field.

The helioseismic travel times, obtained from cross covariance of the
$p$-mode LoS velocities at the observation point and the source
point, show variations of about 1 min along the wave front (the
amplitude depends on inclination $\theta$ of the LoS). Due to the
anisotropy, the travel time averaged along the wave front (like in
the observational procedure) is not zero. Comparison of the
variations of the mean travel times vs. the azimuthal angle of the
observing point shows that the simulation results are in phase with
the observations when Doppler velocities are taken at the level of
300 km above the photosphere (at the same height as the observed
velocities are obtained). The travel time weakly depends on the
height of observations. The amplitude of variations of the travel
times obtained form simulations is about 25\% of the observed
amplitude even for strong fields of 1400--1990 G. It can be an
indication that other effects (for example background flows or
non-uniform distribution of magnetic field) can contribute to the
observed travel time variations. The developed 3D MHD wave
propagation code provides an important tool for further
investigations of local helioseismology in regions with the strong
magnetic field.

\section{Acknowledgements}
We thank to Junwei Zhao for providing us data of variation of the
mean travel times obtained from observations of sunspots.
Calculations were carried out on Columbia supercomputer of NASA Ames
Research Center.

\end{document}